# Two-Way Quantum Communication: 'Secure Quantum Information Exchange'-II. Generalization to Arbitrary Number of Qubits

Ajay K Maurya[1], Manoj K Mishra[1] and Hari Prakash[1, 2]
[1]Physics Department, University of Allahabad, Allahabad, India
[2]Indian Institute of Information Technology, Allahabad, India
Emails: *ajaymaurya.2010@gmail.com*, *manoj.qit@gmail.com*
and *prakash_hari123@rediffmail.com*

**Abstract:**
In this paper, we generalize the secure quantum information exchange (SQIE) protocol, originally proposed by the authors [J. Phys. B: At. Mol. Opt. Phys. **44** (2011) 115504] for secure exchange of one qubit information with each of Alice and Bob, to the case of secure exchange of quantum information of arbitrary qubits with Alice and Bob. We also discuss security of the original and generalized SQIE protocols with respect to the number of qubits with controller, Charlie.

**1. Introduction**

Communication (exchange of information) involves at least a sender to transmit the information and a receiver. Faithful communication occurs only when the receiver is able to reconstruct the exact information that the sender intended to transmit. All the existing practical communication systems, either secure (private) or insecure (public), are capable of transmitting the classical messages (encoded in a string of bits '0' and '1') over a classical channel and are governed by the laws of classical physics.

Over the past decade, researchers have made appreciable progress in the field of quantum information theory and realized that the performance of communication can be enhanced by using transmission channels, which are governed by the laws of quantum mechanics. For example, quantum cryptography [1] allows to distribute a secret-key between two legitimate users say, Alice (sender) and Bob (receiver) with no assumptions of computational powers of eavesdropper, eve. Another example is quantum superdense coding [2] that allows to send two-bit classical message by sending a single two-level particle and sharing an EPR pair, while classically it is required to send a four-level particle. Thus, transmission capacity of classical information transfer is doubled by using EPR-correlation as quantum channel. The two examples given above are the steps towards the transmission of classical information over a quantum channel.

However, if we consider the transfer of quantum information encoded in an unknown qubit, then because of no-cloning theorem, we cannot have many copies of it and therefore, we cannot find the state of unknown qubit. Hence it is not possible to send complete information about the unknown qubit in the form of classical information to the receiver through a classical channel. Also anyhow if state of qubit is known (i.e., $\theta$ & $\phi$ of qubit $\cos(\theta/2)|0\rangle + e^{i\phi}\sin(\theta/2)|1\rangle$ are known), then since $\theta$ & $\phi$ can have infinite possible values that will require infinite number of bits sending to the receiver to construct the qubit. For these reasons, it is not possible to transfer quantum information encoded in a qubit through classical channel.

To overcome such problem, Bennett et al [3] introduced the idea of quantum teleportation (QT) that involves complete transfer of quantum states of a qubit from sender (Alice) to receiver (Bob) using quantum entanglement and restricted amount of classical communication. The idea of QT has been extended from single qubit to multi qubits [4-6] and several schemes have been proposed for experimental realization of QT for photonic states [7], photonic-polarized states [8], optical coherent states [9-10] and atomic states [11-12]. Also several experiments have demonstrated QT with photonic-polarized state [13], quantum state of nucleus [14] and atomic qubits [15-16].

M. Hillery [17] using GHZ state proposed quantum secret sharing in which quantum information splits into two receivers, while Karlsson and Bourennane [18] used GHZ state to teleport single qubit to one of the two receivers, such that only one of them (anyone) can completely reconstruct the qubit depending upon the local measurement result of the other receiver. The use of more than two entangled qubits leads us to the concept of controlled QT in which quantum state can be reconstructed only by one receiver and the local measurement and classical communication by other receiver. Controlled QT is found to be useful in one-way secure quantum networking and in cryptographic conferencing [19-20]. Many authors presented the controlled QT scheme to teleport single qubit information state using GHZ like states [21] and W-state [22]. Further the idea of controlled QT was extended by many authors [23] for teleporting multi-qubit information states.

Very recently, in reference [24], idea of secure quantum information exchange (SQIE) is proposed that enables the faithful exchange of two unknown single qubit states between two legitimate users, Alice and Bob, with the aid of the special kind six-qubit entangled (SSE) state and the classical assistance of a third party Charlie. The SQIE protocol is secure in the sense that either both, Alice and Bob, get their required information states or if this end result is not obtained due to any reason, nobody gets the correct information state. Also Alice and Bob cannot reconstruct the required information states after their measurements and mutual communication without involving Charlie.

More practically, not only exchange of single qubits but also the secure exchange of multi qubits will be required in real world. For this reason, in the present paper, we extend the SQIE protocol from single qubit to multi qubits. Further, we also investigate the security of the original SQIE protocol when the number of qubits with the controller Charlie (the third party) is changed.

## 2. Generalization of SQIE protocol to the information states of arbitrary number of qubits

In this section, we first present a brief review of the original SQIE protocol* [24] and then we will generalize this protocol to achieve secure exchange of information states involving an arbitrary number of qubits between Alice and Bob. Let it be required that Alice has to send arbitrary information state $|\xi\rangle_A^I = [a_0|0\rangle + a_1|1\rangle]_A$ to Bob and Bob has to send another information state $|\eta\rangle_B^I = [b_0|0\rangle + b_1|1\rangle]_B$ to Alice, with the security that either both get their required information states or in case of a failure of this, nobody gets the correct information state. For this purpose, we use the special-kind six-qubit entangled (SSE) states [24],

$$|\psi\rangle_{A_1,B_1,B_2,A_2,C_1,C_2}^E = \tfrac{1}{2}[\sum_{i=0}^{3}|B\rangle_{A_1,B_1}^{(i)} \otimes |B\rangle_{B_2,A_2}^{(i)} \otimes |\phi\rangle_{C_1,C_2}^{(i)}], \tag{1}$$

where, $|B\rangle^{(0,1)} = \tfrac{1}{\sqrt{2}}[|00\rangle \pm |11\rangle]$, $|B\rangle^{(2,3)} = \tfrac{1}{\sqrt{2}}[|01\rangle \pm |10\rangle]$ are the standard bi-partite Bell states and $|\phi\rangle^{(0,1,2,3)}$ are different elements of the set $(|00\rangle, |01\rangle, |10\rangle, |11\rangle)$ taken in any arbitrary order. Subscripts $A_1$, $A_2$ refer to entangled modes with Alice, $B_1$, $B_2$ refer to entangled modes with Bob and $C_1$, $C_2$ refer to entangled modes with the controller Charlie. Superscripts $E$ and $I$ refer to entangled state and information states respectively.

We can write the initial state of composite system as,

$$|\psi\rangle_{A,A_1,B_1,B_2,A_2,C_1,C_2,B} = |\xi\rangle_A^I \otimes |\psi\rangle_{A_1,B_1,B_2,A_2,C_1,C_2}^E \otimes |\eta\rangle_B^I$$
$$= \tfrac{1}{2}[\sum_{i=0}^{3}(|\xi\rangle_A^I \otimes |B\rangle_{A_1,B_1}^{(i)}) \otimes (|B\rangle_{B_2,A_2}^{(i)} \otimes |\eta\rangle_B^I) \otimes |\phi\rangle_{C_1,C_2}^{(i)}]. \tag{2}$$

From Appendix A, we see that,

$$|\xi\rangle_A^I \otimes |B\rangle_{A_1,B_1}^{(i)} = \tfrac{1}{2}[\sum_{r=0}^{3}|B\rangle_{A,A_1}^{(r)} \otimes (\sigma_{B_1}^{(i)}\sigma_{B_1}^{(r)})|\xi\rangle_{B_1}^I],$$
$$|\eta\rangle_B^I \otimes |B\rangle_{B_2,A_2}^{(i)} = \tfrac{1}{2}[\sum_{s=0}^{3}|B\rangle_{B,B_2}^{(s)} \otimes (\sigma_{A_2}^{(i)}\sigma_{A_2}^{(s)})|\eta\rangle_{A_2}^I], \tag{3}$$

where $\sigma^{(i)}$'s are real matrices, $I$, $\sigma_z$, $\sigma_x$, $\sigma_x\sigma_z$ for $i=0, 1, 2, 3$ respectively. Using equations (3), we can write equation (2) as,

---

*The notations used here for original SQIE are not exactly the same as used earlier [24]. The change was required to make the generalization of these results, presented later in this section, more lucid and more presentable.

$$|\psi\rangle_{A,A_1,B_1,B_2,A_2,C_1,C_2,B}$$
$$= \tfrac{1}{8}\left[\sum_{i,r,s=0}^{3}(|B\rangle_{A,A_1}^{(r)}\otimes(\sigma_{B_1}^{(i)}\sigma_{B_1}^{(r)})|\xi\rangle_{B_1}^{I})\otimes(|B\rangle_{B,B_2}^{(s)}\otimes(\sigma_{A_2}^{(i)}\sigma_{A_2}^{(s)})|\eta\rangle_{A_2}^{I})\otimes|\phi\rangle_{C_1,C_2}^{(i)}\right]. \quad (4)$$

Now both, Alice and Bob perform Bell state measurement (BSM) on their qubits $A$, $A_1$ and $B$, $B_2$ respectively, while Charlie measures his qubits $C_1$, $C_2$ in the computational basis $\{|0\rangle, |1\rangle\}$. Alice and Bob convey their BSM results to Charlie through 2-bit classical channels. Charlie, on the basis of these BSM results and his own measurement results, decides about the 2-bit classical information to be conveyed to each of Alice and Bob. On the basis of these classical information, Alice and Bob perform the required unitary transformation on their particles $A_2$ and $B_1$ respectively in order to generate exact replicas of corresponding information states. From equation (4), it is clear that if result of Charlie's measurement is $i$, then Alice performs unitary transformation $(\sigma_{A_2}^{(i)}\sigma_{A_2}^{(s)})^{\dagger}$ and Bob performs unitary transformation $(\sigma_{B_1}^{(i)}\sigma_{B_1}^{(r)})^{\dagger}$ on their particles for the Bob's BSM result $s$ and Alice's BSM result $r$ respectively.

Now we will generalize the SQIE protocol to secure exchange the information states of arbitrary number of qubits between Alice and Bob. Let us consider that Alice wants to send arbitrary $m$-qubit information state, encoded in $m$-qubit modes $\{A\}\equiv(A_1,A_2,...,A_m)$, expressed by

$$|\xi\rangle_{\{A\}}^{I} = [a_0|\tilde{0}\rangle + a_1|\tilde{1}\rangle + ............+a_M|\tilde{M}\rangle]_{\{A\}}, \quad (5)$$

to Bob and Bob wants to send arbitrary $n$-qubit information state, encoded in $n$-qubit modes $\{B\}\equiv(B_1,B_2,...,B_n)$, expressed by

$$|\eta\rangle_{\{B\}}^{I} = [b_0|\tilde{0}\rangle + b_1|\tilde{1}\rangle + ............+b_N|\tilde{N}\rangle]_{\{B\}}, \quad (6)$$

to Alice, with the security that either both get their required information states or, in case of failure of this end result, nobody gets the correct information state. Here, $M\equiv 2^m-1$, $N\equiv 2^n-1$ and for modes $\{A\}$, if $0\leq j\leq M$ and $j=(j_1 j_2 .... j_m)$ in the binary representation, state $|\tilde{j}\rangle_{\{A\}}=|j_1 j_2 .... j_m\rangle_{\{A\}}$. $2^m$-mutually orthogonal states $|\tilde{0}\rangle_{\{A\}}, |\tilde{1}\rangle_{\{A\}}, ...., |\tilde{M}\rangle_{\{A\}}$ form the computational basis for modes $\{A\}$. Similarly for modes $\{B\}$, if $0\leq j\leq N$ and $j=(j_1 j_2 .... j_n)$ in the binary representation, state $|\tilde{j}\rangle_{\{B\}}=|j_1 j_2 .... j_n\rangle_{\{B\}}$. Superscripts $I$ refer to information states.

We now generalize this SQIE protocol. If $p=\max\{m,n\}$, we give $2p$-qubits to Charlie. The problem at this moment is to write the entangled state corresponding to the SSE state of the original SQIE protocol. In the original protocol, Charlie had 2 qubits and SSE state had $2^2=4$ terms. $2p$-qubits of Charlie thus requires $2^{2p}$ terms. If we consider generalized Bell states (GBS) [5] of modes $\{A'\}\equiv(A_1',A_2',....,A_m')$ and $\{B'\}\equiv(B_1',B_2',....,B_m')$ and of $\{B''\}\equiv(B_1'',B_2'',....,B_n'')$ and $\{A''\}\equiv(A_1'',A_2'',....,A_n'')$, there are only $2^{2m}$ and $2^{2n}$ GBS respectively and only one of these gives a family of $2^{2p}$ states. If $m>n$, $2^{2m}=2^{2p}$ but $2^{2n}$ falls shorter than $2^{2p}$ and if $n>m$, $2^{2n}=2^{2p}$ but $2^{2m}$ falls shorter than $2^{2p}$. This problem is circumvented by repeating the members of smaller family of states till $2^{2p}$ states are obtained.

Thus if index $i$ takes values 0, 1, ....., $2^{2p}-1$, we can define indices $i'\equiv i\ (\mathrm{mod}\ 2^{2m})$ and $i''\equiv i\ (\mathrm{mod}\ 2^{2n})$ and write GBS,

$$|E\rangle_{\{A'\},\{B'\}}^{(i)} = |E\rangle_{\{A'\},\{B'\}}^{(i')} \text{ and } |E\rangle_{\{B''\},\{A''\}}^{(i)} = |E\rangle_{\{B''\},\{A''\}}^{(i'')}.$$

The entangled state corresponding to SSE state of the original SQIE protocol can be written as,

$$|\psi\rangle_{\{A'\},\{B'\},\{B''\},\{A''\},\{C\}}^{E} = \frac{1}{2^p}\left[\sum_{i=0}^{2^{2p}-1}|E\rangle_{\{A'\},\{B'\}}^{(i)}\otimes|E\rangle_{\{B''\},\{A''\}}^{(i)}\otimes|\phi\rangle_{\{C\}}^{(i)}\right], \quad (7)$$

Here, modes $\{C\} \equiv (C_1, C_2, ...., C_{2p})$ and states $\{|\phi\rangle_{\{C\}}^{(i)}\}$ are different orthogonal $2^{2p}$-states belonging to the computational basis $(|\tilde{0}\rangle, |\tilde{1}\rangle, ......, |\tilde{P}\rangle)$, $(P \equiv 2^{2p} - 1)$ in $2^{2p}$-dimensional Hilbert space, taken in any order. Superscript $E$ refers to entangled state.

We may now specify the GBS. Since $0 \le i' \le 2^{2m} - 1$, if we express $i'$ in quaternary basis as $i' = (i'_1 i'_2 .... i'_m)$ and write

$$|E\rangle_{\{A'\},\{B'\}}^{(i')} = U_{\{B'\}}^{(i')} |E\rangle_{\{A'\},\{B'\}}^{(0)} = U_{\{B'\}}^{(i')} \frac{1}{2^{m/2}} \sum_{k=0}^{M} |\tilde{k}\rangle_{\{A'\}} \otimes |\tilde{k}\rangle_{\{B'\}}, \tag{8}$$

where

$$U_{\{B'\}}^{(i')} = (\sigma_{B'_1}^{i'_1} \otimes \sigma_{B'_2}^{i'_2} \otimes ......... \otimes \sigma_{B'_m}^{i'_m}), \tag{9}$$

and $\sigma^{(0,1,2,3)} = (I, \sigma_z, \sigma_x, \sigma_x \sigma_z)$. The other GBS can be expressed similarly in the form,

$$|E\rangle_{\{B''\},\{A''\}}^{(i'')} = (U_{\{A''\}}^{(i'')}) |E\rangle_{\{B''\},\{A''\}}^{(i'')} = (U_{\{A''\}}^{(i'')}) \frac{1}{2^{n/2}} \sum_{k=0}^{N} |\tilde{k}\rangle_{\{B''\}} \otimes |\tilde{k}\rangle_{\{A''\}}, \tag{10}$$

where

$$U_{\{A''\}}^{(i'')} = (\sigma_{A''_1}^{i''_1} \otimes \sigma_{A''_2}^{i''_2} \otimes ......... \otimes \sigma_{A''_n}^{i''_n}), \tag{11}$$

$\sigma^{(0,1,2,3)} = (I, \sigma_z, \sigma_x, \sigma_x \sigma_z)$ and $i'' \in (0,1,......,2^{2n} - 1)$ is decimal conversion of quaternary number $(i''_1 i''_2 ....i''_n)$.

Using equations (5), (6) and (7), the initial state of composite system can be written as,

$$|\psi\rangle_{\{A\},\{A'\},\{B'\},\{B''\},\{A''\},\{C\},\{B\}} = |\xi\rangle_{\{A\}}^{I} \otimes |\psi\rangle_{\{A'\},\{B'\},\{B''\},\{A''\},\{C\}}^{E} \otimes |\eta\rangle_{\{B\}}^{I}$$

$$= \frac{1}{2^p} \Big[ \sum_{i=0}^{2^{2p}-1} (|\xi\rangle_{\{A\}}^{I} \otimes |E\rangle_{\{A'\},\{B'\}}^{(i)}) \otimes (|E\rangle_{\{B''\},\{A''\}}^{(i)} \otimes |\eta\rangle_{\{B\}}^{I}) \otimes |\phi\rangle_{\{C\}}^{(i)} \Big]. \tag{12}$$

Qubits in modes $\{A\}, \{A'\}, \{A''\}$ belong to Alice, qubits in modes $\{B\}, \{B'\}, \{B''\}$ belong to Bob and qubits in modes $\{C\}$ belong to Charlie.

From Appendix B, we see that the states, $|\xi\rangle_{\{A\}}^{I} \otimes |E\rangle_{\{A'\},\{B'\}}^{(i)}$ and $|E\rangle_{\{B''\},\{A''\}}^{(i)} \otimes |\eta\rangle_{\{B\}}^{I}$, in equation (12), can be rewritten as,

$$|\xi\rangle_{\{A\}}^{I} \otimes |E\rangle_{\{A'\},\{B'\}}^{(i)} = \frac{1}{2^m} \sum_{r=0}^{2^{2m}-1} |E\rangle_{\{A\},\{A'\}}^{(r)} \otimes U_{\{B'\}}^{(i)} U_{\{B'\}}^{(r)} |\xi\rangle_{\{B'\}}^{I}, \tag{13}$$

$$|E\rangle_{\{B''\},\{A''\}}^{(i)} \otimes |\eta\rangle_{\{B\}}^{I} = \frac{1}{2^n} \sum_{s=0}^{2^{2n}-1} |E\rangle_{\{B\},\{B''\}}^{(s)} \otimes U_{\{A''\}}^{(i)} U_{\{A''\}}^{(s)} |\xi\rangle_{\{A''\}}^{I}, \tag{14}$$

where the GBS $|E\rangle_{\{A\},\{A'\}}^{(r)}$ and $|E\rangle_{\{B\},\{B''\}}^{(s)}$ are given by equations (8) and (10) respectively, unitary matrices $U_{\{B'\}}^{(i)}$, $U_{\{A''\}}^{(i)}$, $U_{\{B'\}}^{(r)}$ and $U_{\{A''\}}^{(s)}$ are given by equations (9), (11), (B.3) and (B.6) respectively.

Using equations (13) and (14), equation (12) can be written as,

$$|\psi\rangle_{\{A\},\{A'\},\{B'\},\{B''\},\{A''\},\{C\},\{B\}}$$

$$= \frac{1}{2^p} \sum_{i=0}^{2^{2p}-1} \Big[ \frac{1}{2^m} \Big\{ \sum_{r=0}^{2^{2m}-1} |E\rangle_{\{A\},\{A'\}}^{(r)} \otimes U_{\{B'\}}^{(i)} U_{\{B'\}}^{(r)} |\xi\rangle_{\{B'\}}^{I} \Big\} \tag{15}$$

$$\otimes \frac{1}{2^n} \Big\{ \sum_{s=0}^{2^{2n}-1} |E\rangle_{\{B\},\{B''\}}^{(s)} \otimes U_{\{A''\}}^{(i)} U_{\{A''\}}^{(s)} |\eta\rangle_{\{A''\}}^{I} \Big\} \otimes |\phi\rangle_{\{C\}}^{(i)} \Big].$$

Now Alice performs generalized $2m$-qubit Bell state measurement (BSM) on her qubits in modes $\{A\}, \{A'\}$ and Bob performs generalized $2n$-qubit BSM on his qubits in modes $\{B\}, \{B''\}$, while Charlie measures his qubits in modes $\{C\}$ in the computational basis $\{|\tilde{0}\rangle, |\tilde{1}\rangle, ......, |\tilde{P}\rangle\}$. Alice and Bob, both, convey their BSM results *r* and *s* to Charlie through 2*m*-bit and 2*n*-bit classical channels respectively. On the basis of these classical information conveyed by Alice and Bob to Charlie and Charlie's measurement result, Charlie can send classical information to Alice and Bob telling them to perform the required unitary transformations on their qubits $\{A''\}$ and $\{B'\}$ respectively, in order to generate exact replicas of the required information states. From equation (15), it is clear that if result of Charlie's measurement is *i*, then Alice performs unitary transformation $(U_{\{A''\}}^{(i)} U_{\{A''\}}^{(s)})^\dagger$ and Bob performs unitary transformation $(U_{\{B'\}}^{(i)} U_{\{B'\}}^{(r)})^\dagger$ on their particles for the Bob's BSM result *s* and Alice's BSM result *r* respectively.

### 3. Security of SQIE protocol with respect to change in the number of qubits going to Charlie

In this section, we discuss dependence of security of SQIE protocol on the number of qubits with the controller Charlie. Let us first consider the case when Charlie has no qubit, *i.e.*, the entangled state shared between Alice and Bob is just a product of the two standard bi-partite Bell states, and we have $|\psi\rangle_{A_1,B_1,A_2,B_2}^E = |B\rangle_{A_1,B_1}^{(i)} \otimes |B\rangle_{A_2,B_2}^{(j)}$, with $i,j \in (0,1,2,3)$. Here, modes $A_1$, $A_2$ are with Alice and modes $B_1$, $B_2$ are with Bob. To complete the SQIE, Alice and Bob, both send their BSM results to Charlie through classical channels. Then Charlie, depending on these results, sends the classical information to Alice and Bob, which are necessary in order to generate the exact replicas of the original information states. Using any process, if Alice and Bob are able to create classical channel between them, then there will be no control of Charlie on the SQIE protocol. By communicating classically, Alice and Bob are able to exchange the information states without any assistance of Charlie. This may lead to a situation when Alice sends her BSM result to Bob but Bob does not send the BSM result to Alice or vice versa, which make the quantum network insecure. Thus, in this case, there is unit probability for insecurity in the quantum network, which is the upper bound.

Let us next we consider the second case when Charlie has single qubit, *i.e.*, the entangled state shared between Alice, Bob and Charlie can be of the form,

$$|\psi\rangle_{A_1,B_1,A_2,B_2,C}^E = \frac{1}{\sqrt{2}}[|B\rangle_{A_1,B_1}^{(i)} \otimes |B\rangle_{A_2,B_2}^{(j)} \otimes |0\rangle_C + |B\rangle_{A_1,B_1}^{(i')} \otimes |B\rangle_{A_2,B_2}^{(j')} \otimes |1\rangle_C],$$

for $i,j,i',j' \in (0,1,2,3)$ with $i \neq i'$ and $j \neq j'$. Modes $A_1$, $A_2$ are with Alice and modes $B_1$, $B_2$ are with Bob, while mode *C* is with Charlie. In this case, Alice and Bob cannot get the required information states by creating classical channel between them, without the assistance of Charlie. The reason is that they do not know which channel ($|B\rangle_{A_1,B_1}^{(i)} \otimes |B\rangle_{A_2,B_2}^{(j)}$ or $|B\rangle_{A_1,B_1}^{(i')} \otimes |B\rangle_{A_2,B_2}^{(j')}$) is setup between them, as it will be determined by Charlie's measurement result; result $|0\rangle_C$ sets the channel $|B\rangle_{A_1,B_1}^{(i)} \otimes |B\rangle_{A_2,B_2}^{(j)}$, while result $|1\rangle_C$ sets the channel $|B\rangle_{A_1,B_1}^{(i')} \otimes |B\rangle_{A_2,B_2}^{(j')}$ between Alice and Bob. However, if Alice and Bob want to ignore the role of Charlie by communicating classically to each other, the probability for getting the required information states successfully is half, *i.e.*, probability for insecurity in the quantum network is half. Thus, the second case is more secure than the first one discussed earlier.

We can now consider the third case, when Charlie has two qubits, which is the original SQIE protocol, introduced by the authors [24]. In this case, if Alice and Bob want to ignore the role of Charlie by creating classical channel between them, there is only one-fourth probability that they are able to get the required information states successfully. The reason is that they do not know which channel ($|B\rangle_{A_1,B_1}^{(0)} \otimes |B\rangle_{A_2,B_2}^{(0)}$ or $|B\rangle_{A_1,B_1}^{(1)} \otimes |B\rangle_{A_2,B_2}^{(1)}$ or $|B\rangle_{A_1,B_1}^{(2)} \otimes |B\rangle_{A_2,B_2}^{(2)}$ or $|B\rangle_{A_1,B_1}^{(3)} \otimes |B\rangle_{A_2,B_2}^{(3)}$) is setup between them, as it will be determined by Charlie's measurement result ($|00\rangle$ or $|01\rangle$ or $|10\rangle$ or $|11\rangle$) respectively. Thus the third case is more secure than the two cases discussed earlier.

If we now increase the number of qubits going to Charlie to three, the entangled state shared between the parties may be of the form, *say*,

$$|\psi\rangle^{E}_{A_1,B_1,A_2,B_2,C_1,C_2,C_3} = \frac{1}{2\sqrt{2}}[|B\rangle^{(0)} \otimes |B\rangle^{(0)} \otimes |000\rangle + |B\rangle^{(1)} \otimes |B\rangle^{(1)} \otimes |001\rangle + |B\rangle^{(2)} \otimes |B\rangle^{(2)} \otimes |010\rangle$$
$$+ |B\rangle^{(3)} \otimes |B\rangle^{(3)} \otimes |011\rangle + |B\rangle^{(0)} \otimes |B\rangle^{(1)} \otimes |100\rangle + |B\rangle^{(1)} \otimes |B\rangle^{(0)} \otimes |101\rangle$$
$$+ |B\rangle^{(2)} \otimes |B\rangle^{(3)} \otimes |110\rangle + |B\rangle^{(3)} \otimes |B\rangle^{(2)} \otimes |111\rangle]_{A_1,B_1,A_2,B_2,C_1,C_2,C_3}.$$

We may involve any eight possible quantum channels between Alice and Bob out of the possible sixteen and Charlie's measurement on his qubits decides the effective channel. Hence, the probability that Alice and Bob are successful in the information exchange without the assistance of Charlie is only one-eighth.

If we increase further the number of qubits going towards Charlie to four, the entangled state shared between them will be of the form, *say*,

$$|\psi\rangle^{E}_{A_1,B_1,A_2,B_2,C_1,C_2,C_3,C_4} = \frac{1}{4}[\sum_{i,j=0}^{3} |B\rangle^{(i)}_{A_1,B_1} \otimes |B\rangle^{(j)}_{A_2,B_2} \otimes |\phi\rangle^{(i)}_{C_1,C_2} \otimes |\phi\rangle^{(j)}_{C_3,C_4}],$$

where $|\phi\rangle^{(i)}$ and $|\phi\rangle^{(j)} \in (|00\rangle, |01\rangle, |10\rangle, |11\rangle)$. In this case, there are sixteen possible quantum channels between Alice and Bob and which one of these sixteen is effective, is decided by Charlie's measurement on his qubits. Hence the probability that Alice and Bob are successful in the information exchange without the assistance of Charlie is only one-sixteenth, *i.e.*, probability for insecurity in the quantum network is only one-sixteenth.

It is clear that the security of the SQIE protocol cannot be increased any further by increasing the number of qubits going towards Charlie beyond four because there are only sixteen possible combinations of product of two standard bi-partite Bell states ($|B\rangle^{(i)}_{A_1,B_1} \otimes |B\rangle^{(j)}_{A_2,B_2}$). Thus if five qubits go to Charlie, the entangled state involves 16 quantum channels and 32 computational basis states of Charlie's qubits. Hence entangled state will have 32 terms and each quantum channel will appear twice. The probability for occurrence of right channel, if Charlie has been sidetracked, is one-sixteenth. Thus one-sixteenth is a lower bound for insecurity in quantum network when Charlie gets four or more qubits.

This consideration can be generalized for exchange of multiple qubits. If Alice and Bob has to send $m$ and $n$ qubit states respectively, the number of possible quantum channels between Alice and Bob is $2^{2(m+n)}$. Thus if Charlie gets $l$ qubits, for $l < 2(m+n)$, the probability for insecurity is $2^{-l}$ and for $l \geq 2(m+n)$, it is $2^{-2(m+n)}$.

## 4. Conclusions

We generalized the original SQIE protocol to exchange the information states of arbitrary number of qubits between two users. We also discussed the security of SQIE protocol and its generalization against the number of qubits with the controller Charlie. We conclude that upper bound probability of insecurity in SQIE protocol is unity and it occurs when the role of Charlie is cut. Also the security of the SQIE protocol cannot be increased indefinitely by increasing the number of qubits going to Charlie. Maximum security is achieved when Charlie receives four qubits as there are four Bell states and there are sixteen possible quantum channels between Alice and Bob. Thus, we find that one-sixteenth is the lower bound for insecurity in the SQIE protocol.

**Acknowledgements**

We are thankful to *Prof. N. Chandra* and *Prof. R. Prakash* for their interest in this work. We would like to thank *Dr. D. K. Singh, Dr. D. K. Mishra, Dr. R. Kumar, Dr. P. Kumar, Mr. Vikram Verma*, and *Mr. Ajay K. Yadav* for helpful and stimulating discussions. One of the authors MKM also acknowledges UGC for financial support under UGC-SRF fellowship scheme and AKM also acknowledges UGC for financial support under UGC D. Phil scholarship scheme.

**Appendix A**

We can write the state $|\xi\rangle_A^I \otimes |B\rangle_{A_1,B_1}^{(i)}$ as,

$$|\xi\rangle_A^I \otimes |B\rangle_{A_1,B_1}^{(i)} = \sum_{r=0}^{3} |B\rangle_{A,A_1}^{(r)} {}^{(r)}\langle B| (|\xi\rangle_A^I \otimes |B\rangle_{A_1,B_1}^{(i)}). \tag{A.1}$$

Since we have $|B\rangle_{A_1,B_1}^{(i)} = \sigma_{B_1}^{(i)} |B\rangle_{A_1,B_1}^{(0)} = \frac{1}{\sqrt{2}} \sigma_{B_1}^{(i)} \sum_{k=0}^{1} |k\rangle_{A_1} \otimes |k\rangle_{B_1}$, where $\sigma_{B_1}^{(i)}$ is real matrix $I, \sigma_z, \sigma_x, \sigma_x\sigma_z$ for $i = 0, 1, 2, 3$ respectively. Then equation (A.1) can be written as,

$$|\xi\rangle_A^I \otimes |B\rangle_{A_1,B_1}^{(i)} = \frac{1}{2} \sum_{r=0}^{3} \sum_{j,k,l=0}^{1} a_j |B\rangle_{A,A_1}^{(r)} {}_A\langle l| \otimes {}_{A_1}\langle l| \sigma_{A_1}^{(r)\dagger} |j\rangle_A \otimes |k\rangle_{A_1} \otimes \sigma_{B_1}^{(i)} |k\rangle_{B_1}$$

$$= \frac{1}{2} \sum_{r=0}^{3} \sum_{j,k=0}^{1} a_j \sigma_{B_1}^{(i)} |B\rangle_{A,A_1}^{(r)} {}_{A_1}\langle j| \sigma_{A_1}^{(r)\dagger} |k\rangle_{A_1} \otimes |k\rangle_{B_1}. \quad \text{(as } {}_A\langle l|j\rangle_A = \delta_{lj}) \tag{A.2}$$

Since, ${}_{A_1}\langle j|\sigma_{A_1}^{(r)\dagger}|k\rangle_{A_1} = {}_{B_1}\langle j|\sigma_{B_1}^{(r)\dagger}|k\rangle_{B_1} = {}_{B_1}\langle k|\sigma_{B_1}^{(r)*}|j\rangle_{B_1} = {}_{B_1}\langle k|\sigma_{B_1}^{(r)}|j\rangle_{B_1}$. Then equation (A.2) becomes,

$$|\xi\rangle_A^I \otimes |B\rangle_{A_1,B_1}^{(i)} = \frac{1}{2} \sum_{r=0}^{3} \sum_{j=0}^{1} a_j \sigma_{B_1}^{(i)} |B\rangle_{A,A_1}^{(r)} \otimes \sigma_{B_1}^{(r)}|j\rangle_{B_1}$$

$$= \frac{1}{2} \sum_{r=0}^{3} |B\rangle_{A,A_1}^{(r)} \otimes \sigma_{B_1}^{(i)} \sigma_{B_1}^{(r)} |\xi\rangle_{B_1}^I. \tag{A.3}$$

Similarly, for the state $|\eta\rangle_B^I \otimes |B\rangle_{B_2,A_2}^{(i)}$, one can write directly using equation (A.3),

$$|\eta\rangle_B^I \otimes |B\rangle_{B_2,A_2}^{(i)} = \frac{1}{2} [\sum_{s=0}^{3} |B\rangle_{B,B_2}^{(s)} \otimes (\sigma_{A_2}^{(i)} \sigma_{A_2}^{(s)}) |\eta\rangle_{A_2}^I]. \tag{A.4}$$

**Appendix B**

We can write the state $|\xi\rangle_{\{A\}}^I \otimes |E\rangle_{\{A'\},\{B'\}}^{(i')}$ as,

$$|\xi\rangle_{\{A\}}^I \otimes |E\rangle_{\{A'\},\{B'\}}^{(i')} = \sum_{r=0}^{2^{2m}-1} |E\rangle_{\{A\},\{A'\}}^{(r)} {}^{(r)}\langle E| (|\xi\rangle_{\{A\}}^I \otimes |E\rangle_{\{A'\},\{B'\}}^{(i')}). \tag{B.1}$$

Using equation (9) and (10), equation (B.1) can be written as,

$$|\xi\rangle_{\{A\}}^I \otimes |E\rangle_{\{A'\},\{B'\}}^{(i')} = \frac{1}{2^m} \sum_{r=0}^{2^{2m}-1} \sum_{j,k,l=0}^{M} a_j U_{\{B'\}}^{(i')} |E\rangle_{\{A\},\{A'\}}^{(r)}$$

$$\{{}_{\{A\}}\langle \tilde{l}| \otimes {}_{\{A'\}}\langle \tilde{l}|(U_{\{A'\}}^{(r)})^\dagger |\tilde{j}\rangle_{\{A\}} \otimes \{|\tilde{k}\rangle_{\{A'\}} \otimes |\tilde{k}\rangle_{\{B'\}}\}$$

$$= \frac{1}{2^m} \sum_{r=0}^{2^{2m}-1} \sum_{j,k=0}^{M} a_j U_{\{B'\}}^{(i')} |E\rangle_{\{A\},\{A'\}}^{(r)} \{{}_{\{A'\}}\langle \tilde{j}|(U_{\{A'\}}^{(r)})^\dagger |\tilde{k}\rangle_{\{A'\}} \otimes |\tilde{k}\rangle_{\{B'\}}\},$$

$$(\text{as } {}_{\{A\}}\langle \tilde{l}|\tilde{j}\rangle_{\{A\}} = \delta_{lj}) \tag{B.2}$$

where

$$U_{\{A'\}}^{(r)} = (\sigma_{A_1'}^{r_1} \otimes \sigma_{A_2'}^{r_2} \otimes \cdots \otimes \sigma_{A_m'}^{r_m}), \tag{B.3}$$

and $r$ is the decimal conversion of quaternary number $(r_1 r_2 \ldots r_m)$. Since

$${}_{\{A'\}}\langle \tilde{j}|(U_{\{A'\}}^{(r)})^\dagger |\tilde{k}\rangle_{\{A'\}} = {}_{\{B'\}}\langle \tilde{j}|(U_{\{B'\}}^{(r)})^\dagger |\tilde{k}\rangle_{\{B'\}} = {}_{\{B'\}}\langle \tilde{k}|(U_{\{B'\}}^{(r)})^* |\tilde{j}\rangle_{\{B'\}} = {}_{\{B'\}}\langle \tilde{k}|(U_{\{B'\}}^{(r)})|\tilde{j}\rangle_{\{B'\}}.$$

Then equation (B.2) becomes,

$$|\xi\rangle_{\{A\}}^I \otimes |E\rangle_{\{A'\},\{B'\}}^{(i')} = \frac{1}{2^m} \sum_{r=0}^{2^{2m}-1} \sum_{j=0}^{M} a_j U_{\{B'\}}^{(i')} |E\rangle_{\{A\},\{A'\}}^{(r)} \otimes (U_{\{B'\}}^{(r)}) |\tilde{j}\rangle_{\{B'\}}$$

$$= \frac{1}{2^m} \sum_{r=0}^{2^{2m}-1} |E\rangle_{\{A\},\{A'\}}^{(r)} \otimes U_{\{B'\}}^{(i')} U_{\{B'\}}^{(r)} |\xi\rangle_{\{B'\}}^I. \tag{B.4}$$

Similarly, for the state $|E\rangle_{\{B''\},\{A''\}}^{(i'')} \otimes |\eta\rangle_{\{B\}}^I$, one can write directly using equation (B.4),

$$|E\rangle^{(i'')}_{\{B''\},\{A''\}} \otimes |\eta\rangle^{I}_{\{B\}} = \frac{1}{2^n} \sum_{s=0}^{2^{2n}-1} |E\rangle^{(s)}_{\{B\},\{B''\}} \otimes U^{(i'')}_{\{A''\}} U^{(s)}_{\{A''\}} |\xi\rangle^{I}_{\{A''\}}, \qquad (B.5)$$

where

$$U^{(i'')}_{\{A''\}} = \sigma^{i''_1}_{A''_1} \otimes \sigma^{i''_2}_{A''_2} \otimes \ldots \otimes \sigma^{i''_n}_{A''_n} \text{ and } U^{(s)}_{\{A''\}} = \sigma^{s_1}_{A''_1} \otimes \sigma^{s_2}_{A''_2} \otimes \ldots \otimes \sigma^{s_n}_{A''_n}, \qquad (B.6)$$

and $s$ is the decimal conversion of quaternary number $(s_1 s_2 \ldots s_n)$.